\documentclass[manuscript]{rmaa}
\usepackage{paralist}
\usepackage{natbib}
\usepackage {graphicx} 
\title     
{
New Analytical Results for Poissonian and  non-Poissonian  Statistics 
of  Cosmic Voids  
}
\author{L. Zaninetti\altaffilmark{1}}

\altaffiltext{1}{Dipartimento di Fisica, Via Pietro Giuria 1,  
10125 Torino,  Italy (zaninetti@ph.unito.it.)}
\shortauthor{Zaninetti}
\shorttitle {Statistics of the voids}
\ReceivedDate{...............} 
\AcceptedDate{...} 
\SetYear{2012}
\resumen{Este documento describe ..
 } 

\abstract
{
Stereology allows
shifting  from
the  3D   distribution
of the volumes  of  Poissonian Voronoi Diagrams
to  their 2D cross-sections.
The  basic  assumption is  that
the  3D statistics of the volumes 
of the voids 
in the local Universe
has a  distribution
function of the gamma-type.
The  standard rule  of conversion  from 3D volumes to
2D circles, 
adopting the standard  rules of stereology, 
produces  a new  probability density
function of the radii which
contains
the Meijer $G$-function.
A  non-Poissonian distribution of volumes is  also
considered.
The distribution of the 3D radii 
of the Sloan Digital Sky Survey
Data Release 7 is best fitted by a 
non-Poissonian distribution in volumes 
as given by the Kiang function with argument
of about two.
}
\addkeyword{statistical distributions }
\addkeyword{galaxies }
\addkeyword{clusters of galaxies} 
\def\aap{A\&A\,  }%% Astronomy and Astrophysics
\def\aj{AJ  }%% The Astronomical Journal
\def\apj{ApJ\,  }%% Astrophysical Journal
\def\apjl{ApJ\,  }%% Astrophysical Journal, Letters
\def\apjs{ApJS  }%% Astrophysical Journal, Supplement
\def\cjaa{Chinese J. Astron. Astrophys.  }% Chinese Journal of Astronomy and Astrophysics
\def\mnras{MNRAS\,  }%% Monthly Notices of the RAS
\def\nat{Nature\,  }%% Nature
\def\pre{Phys. Rev. E   }% % Physical Review E
\def\physa{Phys. A    }% % Physica A
\def\za{Z. Astrophys.  } %Zeitschrift f Astrophysik
\begin{document}
\maketitle

\section{Introduction}

The astronomical analysis  of the  cellular nature  
of the  large
scale structure of our universe 
started with the second CFA2
redshift Survey which produced slices showing that the spatial
distribution of galaxies is not random  but is organized in
filaments which  represent the 2D projection of 3D bubbles, see
\citet{geller}. 
The organization  of astronomical observations
continued  with the 2dF Galaxy Redshift Survey (2dFGRS), 
see
\citet{Colless2001},  
and with the Sloan Digital Sky Survey (SDSS), 
see  \citet{York2000,Abazajian2009}. These catalogs  of slices
allow the determination of the size  of the voids 
as approximated
by circles of a given radius. 
A visual inspection of these slices
allows a rough evaluation of the  largest void, 
which turns to be
$\approx$ 34/h  Mpc. A refined statistics 
requires a digital
version of the radii as given by the catalog of  
cosmic voids  of
SDSS R7, see \citet{Vogeley2011}.

A possible approach  to  the statistics of 
these voids is given by
the Voronoi tessellation, after the two historical papers by
\citet{voronoi_1907,voronoi}. 

Following the nomenclature introduced
by~\citet{okabe}, we call the intersection between a plane and
the 
Poissonian Voronoi
tessellation (PVT)
$V_p(2,3)$. We briefly  recall that the  first
application of the PVT to astrophysics is due to
\citet{kiang}.
         The applications  of Voronoi Diagrams to the galaxies
         started with
         \citet{icke1987},
         where a sequential clustering
         process was adopted in order to insert the initial seeds,
         and continued
         with
         \citet{Weygaert1989,pierre1990,barrow1990,coles1991,
          Weygaert1991a,Weygaert1991b,zaninettig,Ikeuchi1991,
           Subba1992,Weygaert1994,Goldwirth1995,
           Weygaert2002,Weygaert2003,Zaninetti2006}.
         An updated review of 3D Voronoi Diagrams
         applied to  cosmology can be found in
          \citet{Weygaert2002,Weygaert2003}.
         The 3D PVT  can also be
         applied to identify groups of galaxies in
         the structure of a super-cluster,
         see
         \citet{Ebeling1993,Bernardeau1996,Schaap2000,
         Marinoni2002,
         Melnyk2006,
         Schaap2009,
         Elyiv2009}.

A different approach to the intersections 
between bubbles and a
plane is given by stereology, which  is 
the science of the
geometrical relationships 
between structures that exists in three
dimensions (3D)
 and their images,
which are fundamentally two-dimensional (2D).
The absence
of a probability density function (PDF)
for the main parameters of
the PVT area in 2D and the volume in
3D has not allowed the development of
a PDF in 
radii 
of the $V_p(2,3)$ problem.
The publication  with
a  relative test of a new PDF for the cell 
of PVT as
given by
\citet{Ferenc_2007},
allows a  simple parametrization of the cell.
The integral connected with the $V_p(2,3)$ problem
can now be
expressed in analytical terms rather than numerical.
The previous  comments can also  be rewritten 
in the form of some key questions.
\begin{itemize}
\item Is it possible to derive the probability density function
for the  
radii
 of 2D sections in the Poissonian
case? 
\item Is it possible to obtain an analytical expression for
the survival function,
see  Eq.~(\ref{defsurvival}),
of the radii of   2D sections in the
Poissonian case? 
\item Is it possible to derive analytical results
for  the radii of   2D sections in the case of non-Poissonian
seeds or volumes? 
\item Can we apply such obtained
analytical results  to the
 catalog of
cosmic voids as given, for example,
      by the
      SDSS R7?
\end{itemize}

In this paper we
analyze in Section \ref{secadopted} the two 
main PDFs adopted in
order to model the cells of PVT  which are the old
but still widely  used   
Kiang function \citet{kiang} and the  recent 
Ferenc--Neda function \citet{Ferenc_2007}  . 
Section  \ref{secstereology}
reviews 
the
probability of a plane intersecting a given 
sphere , the stereological approach,  
and then insert in the fundamental 
integral of the stereology the 
cell's radius   of  the new PDF.
Section \ref{secobservations}  contains  
the observed statistics 
of 1054 cosmic voids ,  a theoretical comparison  
with the radii of PVT  and 
a comparison of the observed survival function 
of 2dFGRS with our survival function as given by the 
stereology.
An  example   of 
non Poissonian Voronoi
Tessellation (NPVT)
statistics  at the light of the Kiang function is  given in 
Section \ref{secnpvt}.

\section{The distributions adopted for PVT}

\label{secadopted}

We briefly review the PDFs  which  regulate 
the main parameters of  PVTs:
area in 2D, and volume in 3D. 

\subsection{The Kiang function}

The gamma variate $H (x ;c )$  (\citet{kiang})
is
\begin{equation}
 H (x ;c ) = \frac {c} {\Gamma (c)} (cx )^{c-1} \exp(-cx),
\label{kiang}
\end{equation}
where $ 0 \leq x < \infty $, $ c~>0$,
and $\Gamma$ is the gamma function.
The Kiang  PDF has a mean of
\begin{equation}
\mu = 1,
\end{equation}
and variance
\begin{equation}
\sigma^2 = \frac{1}{c}.
\end{equation}
In the case of a 1D PVT,
$c=2$  is an
exact analytical result
and conversely
$c$ is supposed to be 4 or 6
for  2D or  3D  PVTs,
respectively, the so called 
Kiang conjecture  \citet{kiang}.

\subsection{Ferenc--Neda function }

A new PDF
has been recently introduced, \citet{Ferenc_2007},
in order to model the normalized area/volume
in  2D/3D PVT
\begin{equation}
FN(x;d) = C \times x^{\frac {3d-1}{2} } \exp{(-(3d+1)x/2)},
\label{rumeni}
\end{equation}
where $C$ is a constant,
\begin{equation}
C =
\frac
{
\sqrt {2}\sqrt {3\,d+1}
}
{
2\,{2}^{3/2\,d} \left( 3\,d+1 \right) ^{-3/2\,d}\Gamma \left( 3/2\,d+
1/2 \right)
},
\end{equation}
and $d(d=1,2,3)$ is the
dimension of the space under consideration.
We will call this
function the  Ferenc--Neda  PDF;
it has a mean of
\begin{equation}
\mu = 1,
\end{equation}
and variance
\begin{equation}
\sigma^2 = \frac{2}{3d+1}.
\end{equation}
The Ferenc--Neda  PDF  can be obtained from the Kiang function
(\citet{kiang}) by  the transformation
\begin{equation}
c =\frac{3d+1}{2}.
\label{kiangrumeni}
\end{equation}

\subsection{Numerical results}

\label{secnumerical}

In the following, we will
model the PVT
in which the seeds  are computed through a random
process.
The  $\chi^2$ is computed
according to the formula
\begin{equation}
\chi^2 = \sum_{i=1}^N \frac { (T_i - O_i)^2} {T_i},
\label{chisquare}
\end {equation}
where $N$ is the number of bins, 
$T_i$ is the theoretical value,
and $O_i$ is the experimental value.
A first test of the  PDFs
presented in the previous section 
can be made by analyzing the
Voronoi cell normalized area-distribution in 2D
and  normalized volume-distribution in 
3D,
 see
Table~\ref{table_data}.
 \begin{table}
 \caption[]{
The values of $\chi^2$  for
the cell normalized area-distribution function in 2D
and the cell normalized volume-distribution function in 3D
;
here $T_i$ are the  theoretical frequencies and
     $O_i$ are the  sample frequencies.
Here  we have
25 087 Poissonian seeds  in 2D,
21 378 Poissonian seeds  in 3D, and
$40$ intervals in the histogram.
}
 \label{table_data}
 \[
 \begin{array}{llll}
 \hline
dimension &PDF ~& parameters  & \chi^2  \\ \noalign{\smallskip}
 \hline
 \noalign{\smallskip}
2D &  H (x ;c ) (Eq. (\ref{kiang}))   &  c=3.55 & 83.48   \\
\noalign{\smallskip}
\hline
2D & f(x;d) (Eq. (\ref{rumeni}))      &  d=2 & 71.83  \\
 \hline
 \noalign{\smallskip}
3D &  H (x ;c ) ~(Eq. (\ref{kiang}))        &  c=5.53 & 93.86   \\
\noalign{\smallskip}
\hline
3D & f(x;d) ~(Eq. (\ref{rumeni}))           &  d=3 & 134.15  \\
\noalign{\smallskip}
\hline
 \end{array}
 \]
 \end {table}
%siamoqui
In this comparison  $f(x;d)$ 
by \citet{Ferenc_2007},
the number of free parameters is zero   because
$d=2$ or $d=3$ fixes the distribution.
In the case of $H (x ;c )$ by \citet{kiang}, we have 
one free  parameter which is fixed 
by the sample.

\section{Stereology}

\label{secstereology}
We first briefly review   how a
PDF   $f(x)$ changes to
$g(y)$  when a new variable $y (x)$  is introduced. 
We limit  ourselves
to the case in which $y(x)$ is a one-to-one transformation.
The rule for
transforming  a PDF  is
\begin{equation}
g(y) =  \frac  {f(x) } {\vert\frac {dy } {dx} \vert}.
\label{trans} 
\end{equation}

Analytical results  have  shown 
that sections through D-dimensional Voronoi 
tessellations are
not themselves D-1 Voronoi tessellations,
see \citet{Moller1989,Moller1994,Weygaert1996}.
According to \citet{Blower2002}, the
probability of a plane intersecting a given 
sphere is proportional
to the sphere's radius, $R$. 
Cross-sections of radius $r$ may be
obtained from any sphere with a radius greater than or equal to
$r$. We may now write a general expression for 
the probability of
obtaining a cross-section of radius $r$ from the whole
distribution (which is denoted $F(R)$):
\begin{equation}
f(r) = \int_r^{\infty} F(R)  R \frac{1}{R} \frac{r} {\sqrt{R^2
-r^2}} dR,
\label{fundamental}
\end{equation}
which is formula (A7) in \citet{Blower2002}. 
That is to say, $f(r)$
is the probability of finding a bubble of radius $R$,
multiplied
by the probability of intersecting this bubble, 
multiplied by the
probability of obtaining a slice of radius $r$ from this bubble,
integrated over the range of $R\ge r$. 
A first  example is  given
by  the so called monodisperse bubble size distribution (BSD)
which are bubbles of constant radius $R$ and therefore
\begin{equation}
F(R) =\frac{1}{R},
\end{equation}
which is defined in  the interval $[0, R]$
and
\begin{equation}
f(r)= {\frac {r}{\sqrt {{R}^{2}-{r}^{2}}R}},
\end{equation}
which is defined in  the interval $[0, R]$,
see  Eq.~(A4) in \citet{Blower2002}.
The average value  of the radius of the  2D-slices is
\begin{equation}
\overline{r}  =
1/4\,R\pi,
\end{equation}
the variance  is
\begin{equation}
\sigma^2  =
2/3\,{R}^{2}-1/16\,{R}^{2}{\pi }^{2},
\end{equation}
and finally,
\begin{equation}
Skewness = -1.151, \quad
Kurtosis = 0.493.
\end{equation}

\subsection{PVT stereology}

In order to  find our $F(R)$, we now analyze 
the distribution in
effective radius $R$ of the 3D PVT. We assume that 
the volume of
each cell, $v$, is
\begin{equation}
v = \frac{4}{3} \pi R^3.
\end{equation}
In the following, we derive the PDF for
the radius and related
quantities relative to the
 Ferenc--Neda function.
The PDF as a function of the radius 
according to the rule of change of variables (\ref{trans}),
is obtained from
(\ref{rumeni}) on inserting $d=3$:
\begin{equation}
F(R) = {\frac {400 000}{243}}\,{\pi }^{5}{R}^{14}{{\rm e}^{-{\frac
{20}{3}}\, \pi \,{R}^{3}}}.
\label{rumenir}
\end{equation}
The average radius  is
\begin{equation}
\overline{R} = 0.6065,
\end{equation}
and the variance is
\begin{equation}
\sigma^2(R) = 0.00853.
\end{equation}
The  introduction of the scale factor, $b$,
with the new variable
$R=R^{\prime}/b$
transforms Eq.~(\ref{rumenir})  into
\begin{equation}
F(R^{\prime},b) = 
\frac
{
400 000\,{\pi }^{5}{R^{\prime}}^{14}{{\rm e}^{-{\frac {20}{3}}\,{\frac {\pi \,{R^{\prime}
}^{3}}{{b}^{3}}}}}
}
{ 
243\,{b}^{15}
}.
\label{RUMENIRB}
\end{equation}

We now have $F(R)$  as given by Eq.~(\ref{rumenir}) and 
the fundamental integral (\ref{fundamental}), as derived in \citet{Zaninetti2011b},  is 
\begin{eqnarray}
f(r)=&2/3\,{\it K }\,\sqrt [6]{3}\sqrt [3]{10}\sqrt [3]{\pi }r
G^{4, 1}_{3, 5}\left({\frac {100}{9}}\,{\pi }^{2}{r}^{6}\,
\Big\vert\,^{5/6, 1/6, 1/2}_{7/3, 2/3, 1/3, 0, {\frac
{17}{6}}}\right)  \\ 
     & \quad 0 \leq r  \leq 1,
\nonumber
\label{FRMEIJER}
\end{eqnarray}
where ${K}$ is a constant,
\begin{equation}
{K} = 1.6485,
\end {equation}
and  the Meijer $G$-function  is defined
as  in  \citet{Meijer1936,Meijer1941,NIST2010}.
Details  on the real or complex parameters of the 
Meijer $G$-function are  given in the Appendix, \ref{appendixgmeijer}.
Table~\ref{table_parameters} shows the average value, variance, mode, skewness, and kurtosis of the already derived  $f(r)$.
\begin{table}
 \caption[]{
The parameters   of  \lowercase{f(r)}, Eq.~(\ref{FRMEIJER}), relative to
the  $PVT$ case. }
 \label{table_parameters}
 \[
 \begin{array}{ll}
 \hline
Parameter & value   \\ \noalign{\smallskip}
 \hline
 \noalign{\smallskip}
Mean   &  0.4874   \\
\noalign{\smallskip}
\hline
Variance     & 0.02475  \\
\noalign{\smallskip}
\hline
Mode        & 0.553 \\
 \hline
Skewness        & -.5229  \\
 \hline
Kurtosis        &  -.1115  \\
 \hline
 \end{array}
 \]
 \end {table}
Asymptotic series are
\begin{eqnarray}
f(r) \sim
 2.7855\,  r    \\
when \quad \quad r
\rightarrow  0,
\nonumber
\end{eqnarray}
and
\begin{eqnarray}
f(r) \sim - 0.006\, \left( r-1 \right) + 0.136 \, \left( r-1
\right) ^{2}
\\
when \quad r  \rightarrow  1.
\nonumber
\end{eqnarray}
The  distribution function (DF) is
\begin{eqnarray}
DF(r) =   \nonumber  \\
 {\frac {1}{90}}\,{\it K}\,{3}^{5/6}{10}^{2/3} G^{4, 2}_{4,
6}\left({\frac {100}{9}}\,{\pi }^{2}{r}^{6}\, \Big\vert\,^{1, 7/6,
1/2, 5/6}_{8/3, 1, 2/3, 1/3, {\frac {19}{6}}, 0}\right) {\frac
{1}{\sqrt [3]{\pi }}}\\
 \quad 0 \leq r  \leq 1. 
\nonumber  
\end{eqnarray}
The already  defined PDF is  defined in the interval
$0 \leq r  \leq 1$.
In order to make a comparison with a normalized sample
which has a unitarian mean or an 
astronomical sample which  has the mean expressed in 
Mpc, a transformation of scale
should  be introduced.
The  change of variable is $r=x/b$ 
and the resulting PDF is 
\begin{eqnarray}
f(x,b)=  \nonumber  \\
\frac{2}{3}\,{\it K}\,\sqrt [6]{3}\sqrt [3]{10}\sqrt [3]{\pi }x
G^{4, 1}_{3, 5}\left({\frac {100}{9}}\,{\frac {{\pi
}^{2}{x}^{6}}{{b}^{6}}}\, \Big\vert\,^{5/6, 1/6, 1/2}_{7/3, 2/3, 1/3, 0,
{\frac {17}{6}}}\right) (\frac{1}{b})^{2}
\\
 \quad 0 \leq r  \leq b.
\nonumber
\label{FRMEIJERB}
\end{eqnarray}
%newtable  
As an example, Table \ref{table_parameters_twob} 
shows
the statistical parameters for two  different 
values of $b$. Skewness and kurtosis 
do not change with a transformation of scale.
\begin{table}
 \caption[]{
Parameters   of  \lowercase{f(x,b)}, 
Eq.~(\ref{FRMEIJERB}), relative to
the  $PVT$ case. }
 \label{table_parameters_twob}
 \[
 \begin{array}{lll}
 \hline
Parameter ~& b=2.051 & b=34         \\ \noalign{\smallskip}
 \hline
 \noalign{\smallskip}
Mean   &  1.         & 16.57  Mpc   \\
\noalign{\smallskip}
\hline
Variance     & 0.104 & 28.62  Mpc^2 \\
\noalign{\smallskip}
\hline
Mode        & 1.134  &  18.80 Mpc   \\
 \hline
 \end{array}
 \]
 \end {table}

We  briefly  recall  that  a  PDF $ f(x) $ is the first 
derivative of a distribution function (DF) 
$F(x) $ with respect to $x$.
When the DF is unknown  but the PDF known, we have
\begin{equation}
F(x) = \int_0^x f(x) dx.
\end{equation} 
The  survival function (SF)  $S(x)$  is
\begin{equation}
S(x) = 1 -F(x),
\label{defsurvival}
\end{equation}
and  represents the  probability that the variate 
takes a value  greater than $x$.
The  SF  with the
scaling parameter $b$  is
\begin{eqnarray}
SF(x,b)=  \nonumber  \\
1- 0.01831\,{3}^{5/6}{10}^{2/3}
G^{4, 2}_{4, 6}\left({\frac {100}{9}}\,{\frac {{x}^{6}{\pi
}^{2}}{{b}^{6}}}\, \Big\vert\,^{1, 7/6, 1/2, 5/6}_{8/3, 1, 2/3, 1/3, {\frac
{19}{6}}, 0}\right)
{\frac {1}{\sqrt [3]{\pi }}}
\\
 \quad 0 \leq r  \leq b.
\nonumber
\label{sfb}
\end{eqnarray}

\begin{figure*}
\begin{center}
\includegraphics[width=10cm]{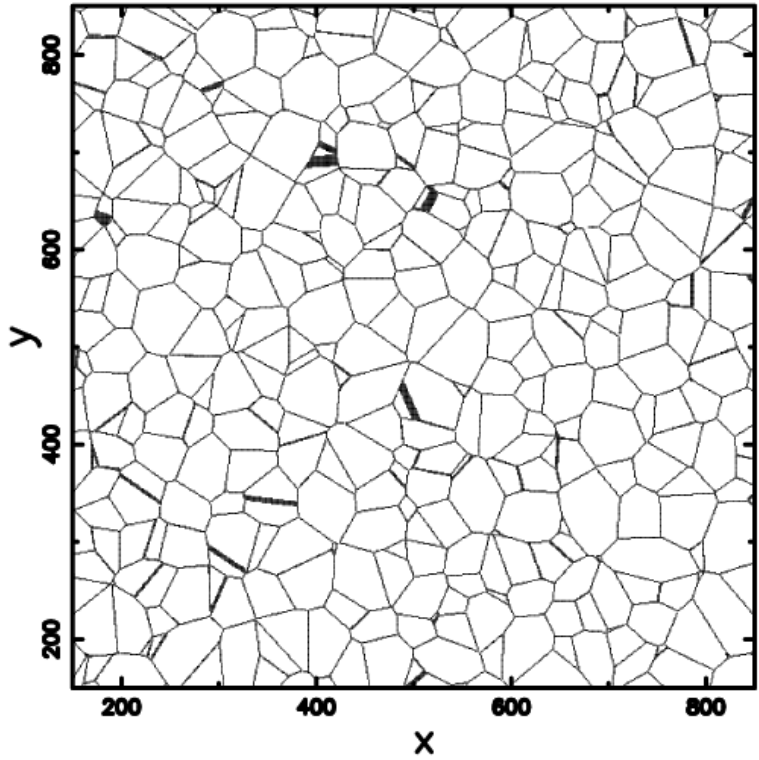}
\end {center}
\caption{
PVT diagram $V_p(2,3)$
when 789  2D cells generated by  15 000 3D seeds are considered.
}
\label{cut_middle}%
    \end{figure*}

\begin{figure*}
\begin{center}
\includegraphics[width=10cm]{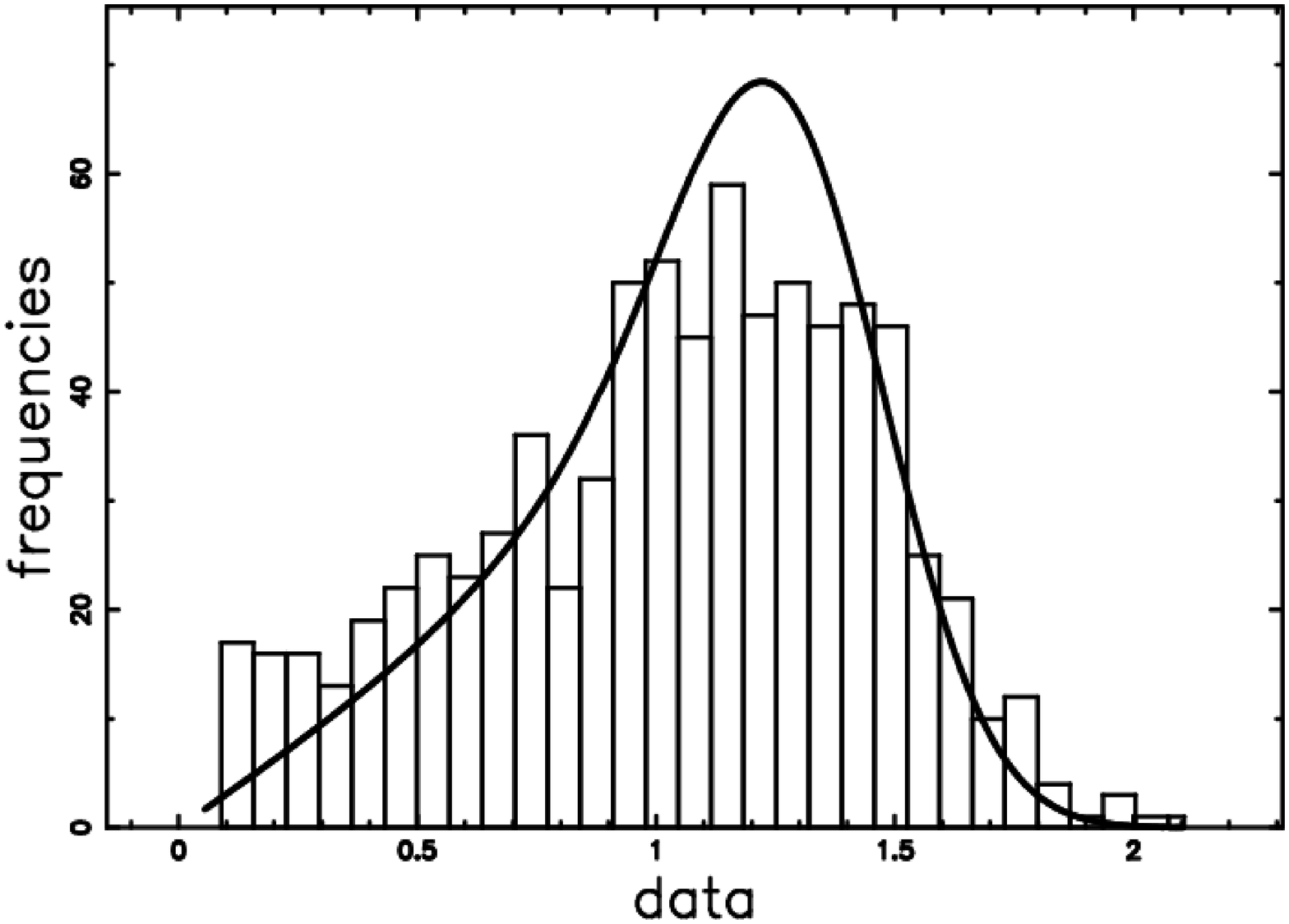}
\end {center}
\caption { Histogram (step-diagram) of PVT  $V_{\lowercase{p}}(2,3)$ when 789 2D
cells, generated by  15 000 3D seeds, are considered. The
superposition of the $f(x,b)$, Eq.~(\ref{FRMEIJERB}), is
displayed. }
\label{frequencies}%
    \end{figure*}

A first  application can be a comparison
between the  real distribution of radii  of
$V_p(2,3)$, see  Fig.~\ref{cut_middle},
and the already obtained
rescaled PDF $f(x,b)$.
%begin figure cut_middle
%end   figure cut_middle
The  fit  with the rescaled $f(x,b)$ is  shown in
Fig.~\ref{frequencies} and
Table \ref{tablechisquare} shows  the $\chi^2$
of three different fitting functions.
%begin figure frequencies
%end   figure frequencies

 \begin{table}
 \caption[]
{
The values of \lowercase {$\chi^2$}  for
the cell normalized area-distribution
of $V_{\lowercase{p}}(2,3)$.
The number of 2D cells is  789,
the 3D seeds are   15 000 and
the number of bins in  the histogram is 30.
}
 \label{tablechisquare}
 \[
 \begin{array}{lll}
 \hline
PDF ~& parameters  & \chi^2  \\ \noalign{\smallskip}
 \hline
 \noalign{\smallskip}
 H (x ;c ) (Eq.~(\ref{kiang}))   &  c=5.8  & 250.8   \\
\noalign{\smallskip}
\hline
f(x;d) (Eq.~(\ref{rumeni}))       &  d=3.53 & 250.8  \\
\noalign{\smallskip}
\hline
f(x,b) (Eq.~(\ref{FRMEIJERB})) & b= 2.0514   &  127 \\
 \hline
 \end{array}
 \]
 \end {table}

The PDF  $f_A$ of the  areas of 
$V_p(2,3)$  can be obtained from 
$f(r)$  by means of the transformation, see \citet{Zaninetti2011b}, 
\begin{equation}
\label{farea}
f_A(A)=f(r) \left ( \left (\frac{A}{\pi}\right )^{1/2}  
\right )\frac{\pi^{-1/2}}{2} {A}^{-1/2},
\end{equation}
\noindent that is, 
\begin{equation}
\label{FAREAG}
f_A(A)=
 0.549\,\sqrt [6]{3}\sqrt [3]{10}
G^{4, 1}_{3, 5}\left({\frac {100}{9}}\,{\frac {{A}^{3}}{\pi }}\,
\Big\vert\,^{5/6, 1/6, 1/2}_{7/3, 2/3, 1/3, 0, {\frac {17}{6}}}\right)
{\pi }^{-2/3}. 
\end{equation}

The already derived  $f_A(A)$ has
average value,  variance, mode, skewness
and kurtosis
as shown in  Table~\ref{table_parameters_area}.
 \begin{table}
 \caption[]{
Parameters   of  {\lowercase{$f}_A(A)$}, Eq.~(\ref{FAREAG}),  relative to
the  $PVT$ case. }
 \label{table_parameters_area}
 \[
 \begin{array}{ll}
 \hline
Parameter ~& value   \\ \noalign{\smallskip}
 \hline
 \noalign{\smallskip}
Mean      &  0.824  \\
\noalign{\smallskip}
\hline
Variance  & 0.204  \\
\noalign{\smallskip}
\hline
Mode      & 0.858 \\
 \hline
Skewness        &0.278    \\
 \hline
Kurtosis        & -0.337  \\
 \hline
 \end{array}
 \]
 \end {table}

\begin{figure*}
\begin{center}
\includegraphics[width=10cm]{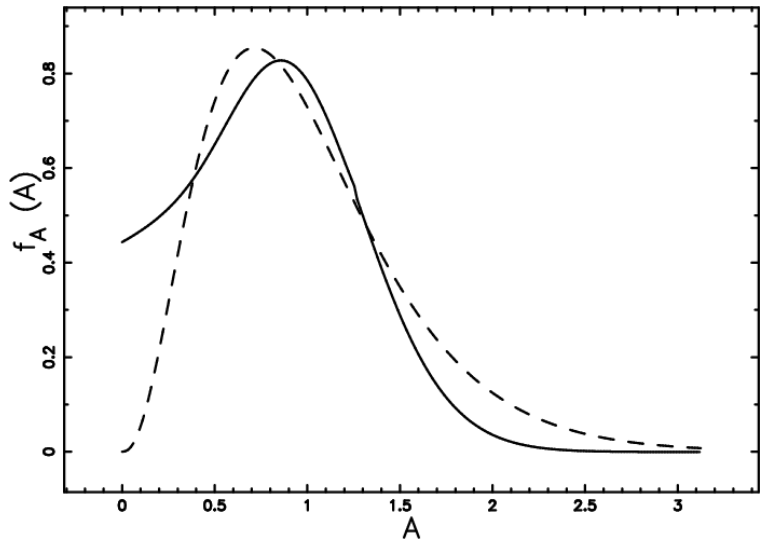}
\end {center}
\caption{
The PDF $f_A$, Eq.~(\ref {FAREAG}), as a function of $A$ (full line) and 
FN(x;d), Eq.~(\ref{rumeni}),  when d=2 
(dotted line).
}
\label{cut_due}
    \end{figure*}

Since, for $r$ close to $0$, $f(r) \sim r$ from Eq.~(\ref{FAREAG}) 
it follows that $f_A(0) \neq 0$, in particular
 $f_A(0)=0.443$   and 
Fig.~\ref{cut_due} shows the graph of $f_A$.  

%begin figure cut_due
%end   figure cut_due
The previous figure shows that sections 
through 3-dimensional Voronoi 
tessellations are
not themselves 2-dimensional  Voronoi tessellations because
$f_A(0)$ has a finite value  rather 
than 0 as does the 2D area distribution;
this fact can be  considered a numerical demonstration 
in agreement with   
\citet{Weygaert1996}.
The distribution function $F_A$ is given by  
\begin{equation}
\label{dfa}
F_A  =
0.018\,{3}^{5/6}{10}^{2/3}
G^{4, 2}_{4, 6}\left({\frac {100}{9}}\,{\frac {{A}^{3}}{\pi }}\,
\Big\vert\,^{1, 7/6, 1/2, 5/6}_{8/3, 1, 2/3, 1/3, {\frac {19}{6}}, 0}\right)
{\frac {1}{\sqrt [3]{\pi }}}.
\end{equation}
Consider a three-dimensional Poisson Voronoi diagram  and suppose 
it intersects a randomly oriented plane $\gamma$:
the resulting cross sections are  polygons.

\begin{figure*}
\begin{center}
\includegraphics[width=10cm]{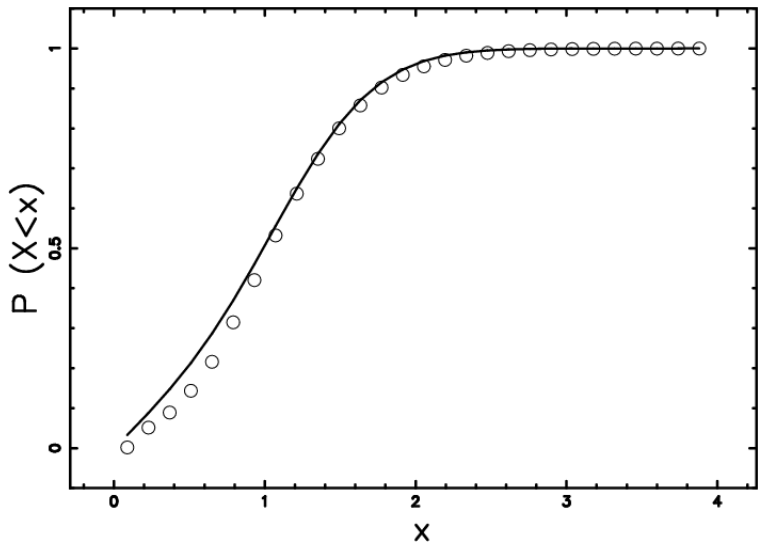}
\end {center}
\caption
{
Comparison between data (empty circles) 
and theoretical curve
(continuous line) 
of  the distribution of areas of the
planar cross  sections.
}
\label{area_xyz}
    \end{figure*}

A  comparison between $F_A$ and 
the area of the irregular polygons 
is shown in Fig.~\ref{area_xyz}. 
In this case the number of seeds is $300 000$ 
and 
we  processed $100168$  irregular polygons
obtained by adding together  results of cuts by  
$41$ triples of  mutually perpendicular  planes.    
The maximum distance  between the two curves 
is $d_{max}=0.039$.  

%begin figure area_xyz
%end   figure area_xyz
As concerns the linear dimension, in our approximation 
the two-dimensional cells were considered 
circles and thus, for consistency, the radius $r$ 
of an irregular  polygon was defined as 
\begin{equation}
r=\left (\frac{A}{\pi} \right )^{1/2},
\label{empradius}
\end{equation}
that is, $r$  is the radius of a circle with the same area, $A$, as the polygon.
The assumption of  sphericity can be considered 
an axiom of the theory here
presented, but for a more realistic 
situation  the stereological 
results will be far more complex.

\section{Statistics  of the voids}
\label{secobservations}

This section first processes 1024  observed cosmic voids 
and then derives the same results from 
the stereological point of view.

\subsection{Observed statistics}

The distribution  of the effective radius  
and 
the radius of the maximal enclosed sphere
between galaxies
of the Sloan Digital Sky Survey
Data Release 7 (SDSS DR7)
has been reported  in
\citet{Vogeley2011}.
This catalog  contains   1054  voids:
Table~\ref{statvoids} shows the basic
statistical  parameters of the 
effective radius,    and  Table~\ref{statvoidsmaximal}, 
the radius of the maximal enclosed sphere.
\begin{table}
 \caption[]
{
The statistical  parameters
of the effective radius in  SDSS DR7.
}
 \label{statvoids}
 \[
 \begin{array}{lc}
 \hline
 \hline
 \noalign{\smallskip}
parameter                  &   value                          \\ \noalign{\smallskip}
elements                   &  1024               \\ \noalign{\smallskip}
mean                       &  18.23h^{-1}~ Mpc   \\ \noalign{\smallskip}
variance                   &  23.32h^{-2}~ Mpc^2 \\ \noalign{\smallskip}
standard~ deviation        &  4.82h^{-1} ~ Mpc   \\ \noalign{\smallskip}
skewness                   &  0.51         \\ \noalign{\smallskip}
kurtosis                   &  0.038        \\ \noalign{\smallskip}
maximum ~value             &  34.12h^{-1}~ Mpc   \\ \noalign{\smallskip}
minimum ~value             &  9.9h^{-1}~   Mpc   \\ \noalign{\smallskip} \hline
 \hline
 \end{array}
 \]
 \end {table}

\begin{table}
 \caption[]
{
The statistical  parameters
of 
the radius of the maximal enclosed sphere
in  SDSS DR7.
}
 \label{statvoidsmaximal}
 \[
 \begin{array}{lc}
 \hline
 \hline
 \noalign{\smallskip}
parameter                  &   value                          \\ \noalign{\smallskip}
elements                   &  1054              \\ \noalign{\smallskip}
mean                       &  12.95h^{-1}~ Mpc   \\ \noalign{\smallskip}
variance                   &  6.99 ^{-2}~ Mpc^2 \\ \noalign{\smallskip}
standard~ deviation        &  2.64 h^{-1} ~ Mpc   \\ \noalign{\smallskip}
skewness                   &  1.47         \\ \noalign{\smallskip}
kurtosis                   &  2.401        \\ \noalign{\smallskip}
maximum ~value             &  25.69h^{-1}~ Mpc   \\ \noalign{\smallskip}
minimum ~value             &  10 h^{-1}~   Mpc   \\ \noalign{\smallskip} \hline
 \hline
 \end{array}
 \]
 \end {table}

\subsection{PVT statistics}

\begin{figure}
\begin{center}
\includegraphics[width=10cm]{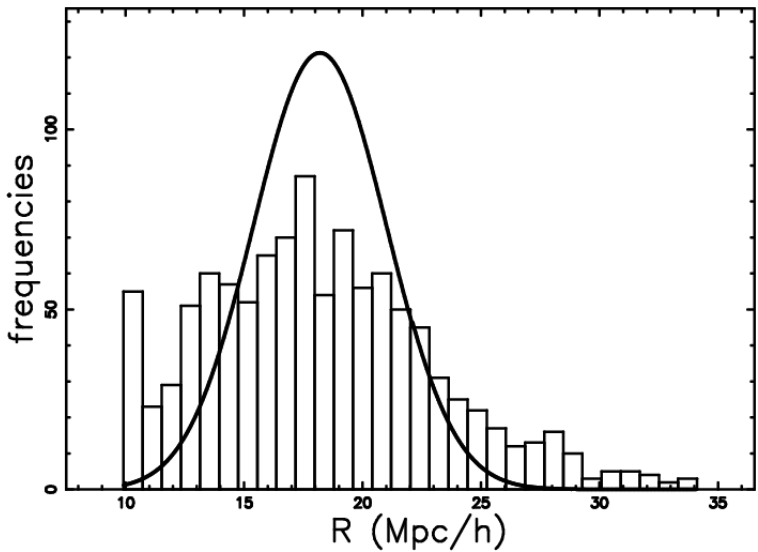}
\end {center}
\caption
{
Histogram (step-diagram)  of
the
effective radius in  SDSS DR7
with a superposition of the
PDF  in radius of the PVT spheres, $F(R,b)$,
 as represented by Eq.~(\ref{RUMENIRB}).
The number of bins is 30, and b = 30.05 Mpc.
}
\label{stat_rumeni3d_sdss}
    \end{figure}

Fig.~\ref{stat_rumeni3d_sdss} shows  a superposition 
of the effective
radius of the voids in the SDSS DR7 with a 
the curve of
the theoretical PDF in the radii, $F(R,b)$, as  given by
Eq.~(\ref{RUMENIRB}).  
Table~\ref{stattheoretical} shows
the theoretical  statistical  parameters.
%begin figure stat_rumeni3d_sdss
% end figure stat_rumeni3d_sdss

\begin{table}
 \caption[]
{
The statistical  parameters
of the theoretical radius of the  PVT spheres 
as represented by Eq.~(\ref{RUMENIRB})
when  {\lowercase{b}}=30.05 M\lowercase{pc}.
}
 \label{stattheoretical}
 \[
 \begin{array}{lc}
 \hline
 \hline
 \noalign{\smallskip}
parameter                  &   value                          \\ \noalign{\smallskip}
mean                       &  18.23h^{-1}~ Mpc   \\ \noalign{\smallskip}
variance                   &   7.70h^{-2}~ Mpc^2 \\ \noalign{\smallskip}
standard~ deviation        &   2.77h^{-1} ~ Mpc   \\ \noalign{\smallskip}
mode                       &   18.49h^{-1}~ Mpc   \\ \noalign{\smallskip}
skewness                   &   0.0142         \\ \noalign{\smallskip}
kurtosis                   &  -0.0514         \\ \noalign{\smallskip}
 \hline
 \end{array}
 \]
 \end {table}

Table \ref{table_chi2sdss}  shows the values of $\chi^2$
for the   main  PDFs here considered.
 \begin{table}
 \caption[]{
Values of  {\lowercase{$\chi^2$}}  for the effective radius in  SDSS DR7 for
different distributions  when the number of bins is 30. 
In this
comparison, the averaged value of the astronomical radii is one. }
 \label{table_chi2sdss}
 \[
 \begin{array}{lll}
 \hline
PDF ~& parameters  & \chi^2  \\ \noalign{\smallskip}
 \hline
 \noalign{\smallskip}
 H (x ;c ), (Eq. (\ref{kiang}))   &  c=14.24  & 53   \\
\noalign{\smallskip}
\hline
f(x;d), (Eq. (\ref{rumeni}))       &  d=9.1    & 53  \\
\noalign{\smallskip}
\hline
f(x,b), (Eq. (\ref{FRMEIJERB}))   &  b=2.051  & 182 \\
 \hline
F(R,b), (Eq. (\ref{RUMENIRB}))    &  b=1.648  & 407 \\
 \hline
F_K(R,b,c), (Eq. (\ref{KIANGVARC}) & b=31.33 ~c=1.76  & 66.121 \\
 \hline
 \end{array}
 \]
 \end {table}
The statistics  of the voids
can also be visualized
through  the SF, see an application to the 2dFGRS as given by
\citet{Patiri2006,Benda-Beckmann2008}.

\begin{figure}
\begin{center}
\includegraphics[width=10cm]{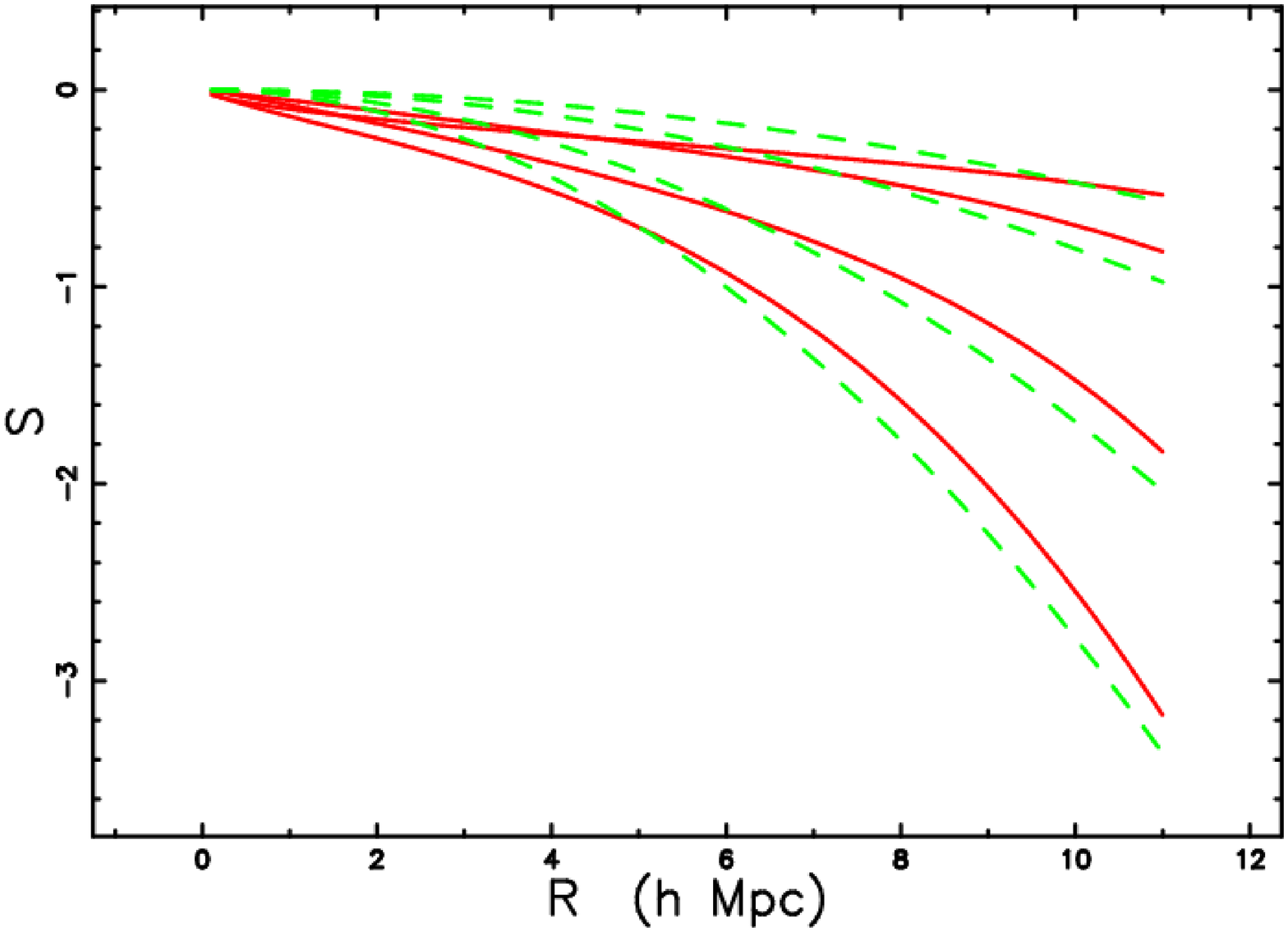}
\end {center}
\caption{
The survival function,
$S_{SS}$,
for the
self-similar distribution in radius of $N/S1$,
$N/S2$, $N/S3$ and $N/S4$,
as reported in Fig.~4 of
\citet{Benda-Beckmann2008} (full red lines),
as represented by (\ref{survivalss}).
The survival function, $ SF(x,b) $,
of the radius of the  distribution
which involves the Meijer $G$-function
for $V_p(2,3)$
as represented by
(\ref{sfb})
when $b= 12 $ Mpc,
     $b= 14 $ Mpc,
     $b= 17 $ Mpc, and
     $b= 19 $ Mpc
(dashed green  lines).
}
 \label{comparison_sample}%
 \end{figure}

The statistics of the voids
between galaxies have been also analysed
in \citet{Benda-Beckmann2008} 
with the following self-similar
SF in the following, $S_{SS}$,
\begin{equation}
S_{SS}=
{{\rm e}^{- \left( {\frac {R}{s_{{1}}\lambda}} \right) ^{p_{{1}}}-
 \left( {\frac {R}{s_{{2}}\lambda}} \right) ^{p_{{2}}}}},
\label{survivalss}
\end{equation}
where $\lambda$ is the mean separation between galaxies,
$s_1$ and $s_2$ are two length factors,
and  $p_1$ and $p_2$ two powers.
A final comparison between the four samples
of void size statistics as
represented in Fig. 4 of
\citet{Benda-Beckmann2008} 
and our
survival function
of the radius 
for $V_p(2,3)$ as  given by 
Eq.~(\ref{sfb})
is shown in Fig.~\ref{comparison_sample}.

%begin figure comparison_sample
% end figure comparison_sample
More details as well the PDF of the self-similar 
distribution  can be found  in 
\citet{Zaninetti2010g}.  

\section{NPVT statistics}

\label{secnpvt}

\begin{figure}
\begin{center}
\includegraphics[width=10cm]{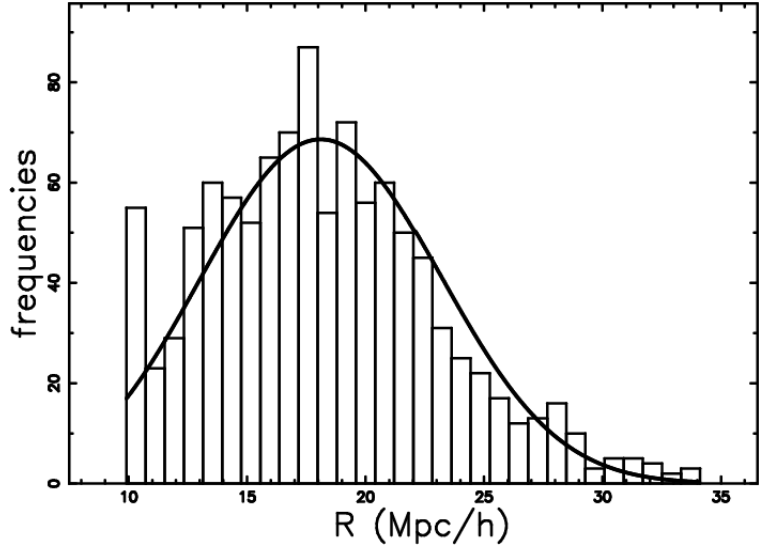}
\end {center}
\caption
{
Histogram (step-diagram)  of
the
effective radius in the  SDSS DR7
with a superposition of the
PDF  in radius of the NPVT spheres, $F_K(R,b,c)$,
 as represented by Eq.~(\ref{KIANGVARC}).
The number of bins is 30, $b$=31.33 Mpc,
and  $c=1.768$. 
}
\label{stat_kiangvarc_sdss}
    \end{figure}

An   example  of non NPVT is
represented  by a distribution  in volume
which  follows a Kiang function as given
by Eq.~(\ref{kiang}).
The case  of PVT volumes indicates $c=5$,
see Eq.~(\ref{kiangrumeni}),  
or  $c=6$, 
the so called Kiang conjecture:
we will  take  $c$ as a variable.
The resulting  distribution in radius  
once  the scaling parameter $b$ is introduced 
is  
\begin{equation}
F_K(R,b,c) = 
\frac
{
4\,c \left( 4/3\,{\frac {c\pi \,{R}^{3}}{{b}^{3}}} \right) ^{c-1}{
{\rm e}^{-4/3\,{\frac {c\pi \,{R}^{3}}{{b}^{3}}}}}\pi \,{R}^{2}
}
{ 
\Gamma  \left( c \right) {b}^{3}
}.
\label{KIANGVARC}
\end{equation}
The average radius  is
\begin{equation}
\overline{R} = 
\frac
{
\sqrt [3]{2}\sqrt [3]{3}b\Gamma  \left( 1/3+c \right)
}
{
2\,\sqrt [3]{c}\sqrt [3]{\pi }\Gamma  \left( c \right)
},
\end{equation}
and the variance is
\begin{equation}
\sigma^2(R) =
\frac   
{
-{3}^{2/3}{2}^{2/3}{b}^{2} \left( -\Gamma  \left( 2/3+c \right) 
\Gamma  \left( c \right) + \left( \Gamma  \left( 1/3+c \right) 
 \right) ^{2} \right) 
}
{
4\,{c}^{2/3}{\pi }^{2/3} \left( \Gamma  \left( c \right)  \right) ^{2}
}.
\end{equation}
The skewness  is
\begin{equation}
\gamma =
\frac
{
 \left( \Gamma  \left( c \right)  \right) ^{3}c-3\,\Gamma  \left( c
 \right) \Gamma  \left( 1/3+c \right) \Gamma  \left( 2/3+c \right) +2
\, \left( \Gamma  \left( 1/3+c \right)  \right) ^{3}
}
{
 \left( \Gamma  \left( 2/3+c \right) \Gamma  \left( c \right) -
 \left( \Gamma  \left( 1/3+c \right)  \right) ^{2} \right) ^{3/2}
},
\end{equation}
and the  
kurtosis is  given by a complicated analytical expression.
Fig.~\ref{stat_kiangvarc_sdss} shows  a superposition 
of   the effective
radii of the voids in  SDSS DR7 with a 
superposition  of the curve of
the theoretical PDF in the radius, $F_K(R,b,c) $, 
as  represented by
Eq.~(\ref{KIANGVARC}).
Table \ref{table_chi2sdss}  shows the values of $\chi^2$
.
Table~\ref{statkiangvarc} shows
the theoretical  statistical  parameters.
%begin figure stat_kiangvarc_sdss
% end figure stat_kiangvarc_sdss

\begin{table}
 \caption[]
{
The statistical  parameters
of the theoretical radius of the  NPVT spheres 
as represented by Eq.~(\ref{KIANGVARC})
when ${\lowercase{b}}$=31.33 M\lowercase{pc}
and  ${\lowercase{c}}=1.768$.
}
 \label{statkiangvarc}
 \[
 \begin{array}{lc}
 \hline
 \hline
 \noalign{\smallskip}
parameter                  &   value                          \\ \noalign{\smallskip}
mean                       &  18.23h^{-1}~ Mpc   \\ \noalign{\smallskip}
variance                   &   23.31h^{-2}~ Mpc^2 \\ \noalign{\smallskip}
standard~ deviation        &   4.82h^{-1} ~ Mpc   \\ \noalign{\smallskip}
skewness                   &   0.072         \\ \noalign{\smallskip}
kurtosis                   &  -0.162         \\ \noalign{\smallskip}
 \hline
 \end{array}
 \]
 \end {table}
The result of the integration of the fundamental Eq.~(\ref{fundamental}) inserting $c$=2 gives 
the following PDF for the radius of
the cuts
\begin{eqnarray}
f(r)_{NPVTK} = 
 3.4148\,\sqrt [6]{3}\sqrt [3]{\pi }r{2}^{2/3}
G^{4, 0}_{2, 4}\left({\frac {16}{9}}\,{\pi }^{2}{r}^{6}\, \Big\vert\,^{1/6,
1/2}_{4/3, 2/3, 1/3, 0}\right)
 \\
 \quad 0 \leq r  \leq 1. \label{FRMEIJERNPVTK}
\nonumber
\end{eqnarray}
The statistics of NPVT cuts with $c$=2  are shown in
Table~\ref{table_parametersnpvtk}.
 \begin{table}
 \caption[]{
NPVT parameters   of 
\lowercase{$f(r)}_{NPVTK}$, Eq.~(\ref{FRMEIJERNPVTK}). }
 \label{table_parametersnpvtk}
 \[
 \begin{array}{ll}
 \hline
Parameter ~& value   \\ \noalign{\smallskip}
 \hline
 \noalign{\smallskip}
Mean   &  0.488   \\
\noalign{\smallskip} \hline
Variance     & 0.0323 \\
\noalign{\smallskip} \hline
Mode        & 0.517 \\
 \hline
Skewness        & -.114  \\
 \hline
Kurtosis        &  2.614  \\
 \hline
 \end{array}
 \]
 \end {table}

\begin{figure}
\begin{center}
\includegraphics[width=10cm]{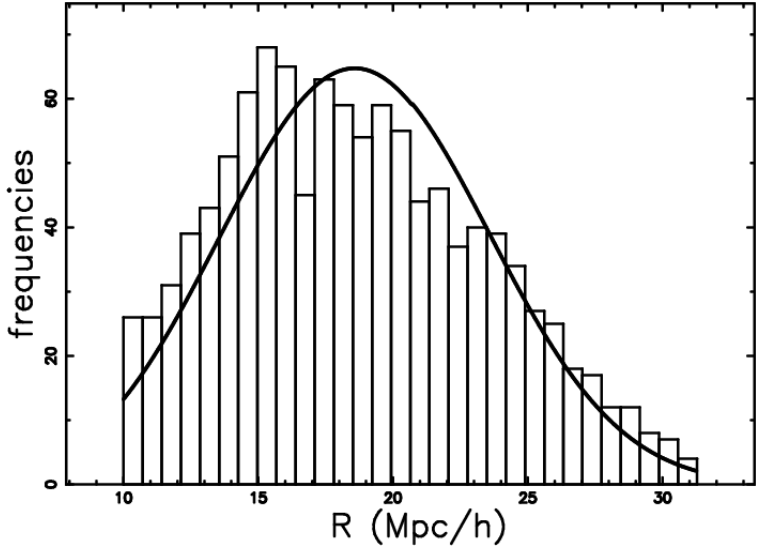}
\end {center}
\caption
{
Histogram (step-diagram)  of
the
simulated  effective radius of  SDSS DR7
with a superposition of the
PDF  in radius of the PVT spheres
as represented by Eq.~(\ref{KIANGVARC}).
The artificial sample has a minimum value   
of  10/h  Mpc,
the number of bins is 30, $b$= 31.5/h Mpc, and $c=1.3$.
}
\label{stat_cvar_gene}
    \end{figure}

On introducing the scaling parameter $b$, the  
PDF which describes
the radius of the cut becomes
\begin{eqnarray}
f(x,b)_{NPVTK} = 
3.4148\,\sqrt [6]{3}\sqrt [3]{\pi }x{2}^{2/3}
G^{4, 0}_{2, 4}\left({\frac {16}{9}}\,{\frac {{\pi }^{2}{x}^{6}}{{b}^{6}}}\,
\Big\vert\,^{1/6, 1/2}_{4/3, 2/3, 1/3, 0}\right)
{b}^{-2}
 \\
 \quad 0 \leq r  \leq b. \label{frmeijerbnpvtk}
\nonumber
\end{eqnarray}
The  SF  of the {\it second} 
NPVT case, $SF_{NPVTK}$, with the
scaling parameter $b$, is
\begin{eqnarray}
SF(x,b)_{NPVTK}=   
1- 0.2845\,{3}^{5/6}\sqrt [3]{2}
G^{4, 1}_{3, 5}\left({\frac {16}{9}}\,{\frac {{\pi }^{2}{x}^{6}}{{b}^{6}}}\,
\Big\vert\,^{1, 1/2, 5/6}_{5/3, 1, 2/3, 1/3, 0}\right)
{\frac {1}{\sqrt [3]{\pi }}}
\\
 \quad 0 \leq r  \leq b.
 \nonumber
\label{sfbnpvtk}
\end{eqnarray}
A careful  exploration of the distribution  
in effective radius
 of
SDSS DR7  reveals that the detected  voids have  radius
$\geq $ 10/h Mpc.
This observational fact demands the  
generation of random numbers
in  the distribution in radii of the 3D cells as
given by Eq.~(\ref{KIANGVARC})   with a minimal value  of
10/h  Mpc.
The  artificial sample is generated
through a numerical computation
of the inverse function~\citet{Brandt1998} and  displayed 
in Fig.~\ref{stat_cvar_gene};  the sample's  statistics   
are shown
in Table~\ref{statsimulatedcvar}.

%begin figure stat_cvar_gene
% end figure stat_cvar_gene

\begin{table}
 \caption[]
{
The statistical  parameters
of the artificially generated   radius 
with a lower bound
of   10 /\lowercase{h}  M\lowercase{pc}, \lowercase{$c=1.3$} 
and \lowercase{$b$} =31.5/\lowercase{h}  M\lowercase{pc}.
}
 \label{statsimulatedcvar}
 \[
 \begin{array}{lc}
 \hline
 \hline
 \noalign{\smallskip}
parameter                  &   value                          \\ \noalign{\smallskip}
mean                       & 18.69  h^{-1}~ Mpc   \\ \noalign{\smallskip}
variance                   & 22.74  h^{-2}~ Mpc^2 \\ \noalign{\smallskip}
standard~ deviation        & 4.76 h^{-1} ~ Mpc   \\ \noalign{\smallskip}
skewness                   & 0.33           \\ \noalign{\smallskip}
kurtosis                   & -0.623               \\ \noalign{\smallskip}
maximum ~value             & 31.27  h^{-1}~ Mpc   \\ \noalign{\smallskip}
minimum ~value             & 10    h^{-1}~   Mpc   \\ \noalign{\smallskip} \hline
 \hline
 \end{array}
 \]
 \end {table}
%tabella1

\section{Conclusions}

{\bf  PVT Statistics} The  approach as  given  
by  the  stereology  to
the PDF in radii of the  circles which  result  from  the
intersection between a plane and a randomly disposed 
spheres of
radius $R$   is  actually limited  to the case of 
mono-disperse
spheres  of radius $R$ and to  a power law 
with radius  $\propto
R^{-\alpha}$  \citet{Blower2002}. 
Here  adopting the 
same type of
demonstration   we  simply substitute into formula
(\ref{fundamental}) a new distribution for  the 
generalized radii, $(R)$,  
of PVT. 
The resulting distribution in radii,
$(r)$,
  of the circles
of intersection  involves the Meijer $G$-function. 
A first test
on this new PDF for the radii  was 
performed on the
2dFGRS  catalog 
and the theoretical  $V_p(2,3)$  cells
were compared  with  other fitting  functions,
see Fig.~\ref{comparison_sample}.

{\bf NPVT Statistics}

Among the infinite number of 3D seeds which are 
non-Poissonian,  we selected 
 a distribution  in volume
which  follows a Kiang function as given
by Eq. (\ref{kiang}) with $c\approx$2.
 
A careful comparison with the  measured 
effective radii
permits us to say that the  
NPVT case  here considered  
is a good  model  
because it  can reproduce  the 3D average radius 
and the variance, see Table \ref{statkiangvarc}. 
The model for the effective  radius of the voids 
as given  by the Kiang distribution in volumes
with $c$ variable can also be used to generate an
artificial sample   of the effective radius of the voids,
see Fig.~\ref{stat_cvar_gene} and Table \ref{statsimulatedcvar}.

\section*{ Acknowledgements}
I would like to thank the anonymous referee for constructive
comments on the text and Mario Ferraro for 
positive discussions  on the Voronoi Diagrams.

\hrulefill
%\bibliography{biblio}

\begin{thebibliography}
\expandafter\ifx\csname natexlab\endcsname\relax\def\natexlab#1{#1}\fi
\expandafter\ifx\csname href\endcsname\relax
  \def\href#1#2{}\fi
\expandafter\ifx\csname urllinklabel\endcsname\relax
  \def\urllinklabel{[LINK]}\fi
\expandafter\ifx\csname adsurllinklabel\endcsname\relax
  \def\adsurllinklabel{[ADS]}\fi

\bibitem[{{Abazajian} {et~al.}(2009){Abazajian}, {Adelman-McCarthy},
  {Ag{\"u}eros}, {Allam}, {Allende Prieto}, \& et~al.}]{Abazajian2009}
{Abazajian}, K.~N., {Adelman-McCarthy}, J.~K., {Ag{\"u}eros}, M.~A., {Allam},
  S.~S., {Allende Prieto}, C., \& et~al. 2009, \apjs, 182, 543


\bibitem[{{Barrow} \& {Coles}(1990)}]{barrow1990}
{Barrow}, J.~D. \& {Coles}, P. 1990, \mnras, 244, 188


\bibitem[{{Bernardeau} \& {van de Weygaert}(1996)}]{Bernardeau1996}
{Bernardeau}, F. \& {van de Weygaert}, R. 1996, \mnras, 279, 693


\bibitem[{Blower {et~al.}(2002)Blower, Keating, Mader, \&
  Phillips}]{Blower2002}
Blower, J., Keating, J., Mader, H., \& Phillips, J. 2002, Journal of
  Volcanology and Geothermal Research, 120, 1


\bibitem[{{Brandt} \& {Gowan}(1998)}]{Brandt1998}
{Brandt}, S. \& {Gowan}, G. 1998, Data Analysis: Statistical and Computational
  Methods for Scientists and Engineers (New York: Springer-Verlag).


\bibitem[{{{Bratley}, {P.} and {Fox}, B. L.}(1988)}]{Bratley1988}
{{Bratley}, {P.} and {Fox}, B. L.} 1988, ACM Trans. Math. Softw., 14, 88


\bibitem[{Chiu {et~al.}(1996)Chiu, Weygaert, \& Stoyan}]{Weygaert1996}
Chiu, S.~N., Weygaert, R. V.~D., \& Stoyan, D. 1996, Adv. in Appl.
  Prob., 28, 356


\bibitem[{{Coles}(1991)}]{coles1991}
{Coles}, P. 1991, \nat, 349, 288


\bibitem[{{Colless} {et~al.}(2001){Colless}, {Dalton}, {Maddox}, \& {et
  al.}}]{Colless2001}
{Colless}, M., {Dalton}, G., {Maddox}, S., \& {et al.} 2001, \mnras, 328, 1039


\bibitem[{{Ebeling} \& {Wiedenmann}(1993)}]{Ebeling1993}
{Ebeling}, H. \& {Wiedenmann}, G. 1993, \pre, 47, 704


\bibitem[{{Elyiv} {et~al.}(2009){Elyiv}, {Melnyk}, \& {Vavilova}}]{Elyiv2009}
{Elyiv}, A., {Melnyk}, O., \& {Vavilova}, I. 2009, \mnras, 394, 1409


\bibitem[{{Ferenc} \& {N{\'e}da}(2007)}]{Ferenc_2007}
{Ferenc}, J.-S. \& {N{\'e}da}, Z. 2007, \physa, 385, 518


\bibitem[{{Ferraro} \& {Zaninetti}(2011)}]{Zaninetti2011b}
{Ferraro}, M. \& {Zaninetti}, L. 2011, \pre, 84, 041107


\bibitem[{{Geller} \& {Huchra}(1989)}]{geller}
{Geller}, M.~J. \& {Huchra}, J.~P. 1989, Science, 246, 897


\bibitem[{{Goldwirth} {et~al.}(1995){Goldwirth}, {da Costa}, \& {van de
  Weygaert}}]{Goldwirth1995}
{Goldwirth}, D.~S., {da Costa}, L.~N., \& {van de Weygaert}, R. 1995, \mnras,
  275, 1185


\bibitem[{{Ho} {et~al.}(2007){Ho}, {James}, \& {Lau}}]{H02007}
{Ho}, M.-W., {James}, L.~F., \& {Lau}, J.~W. 2007, ArXiv e-prints


\bibitem[{{Icke} \& {van de Weygaert }(1987)}]{icke1987}
{Icke}, V. \& {van de Weygaert }, R. 1987, \aap, 184, 16


\bibitem[{{Ikeuchi} \& {Turner}(1991)}]{Ikeuchi1991}
{Ikeuchi}, S. \& {Turner}, E.~L. 1991, \mnras, 250, 519


\bibitem[{{Kiang}(1966)}]{kiang}
{Kiang}, T. 1966, \za, 64, 433


\bibitem[{{Marinoni} {et~al.}(2002){Marinoni}, {Davis}, {Newman}, \&
  {Coil}}]{Marinoni2002}
{Marinoni}, C., {Davis}, M., {Newman}, J.~A., \& {Coil}, A.~L. 2002, \apj, 580,
  122


\bibitem[{Meijer(1936)}]{Meijer1936}
Meijer, C. 1936, Nieuw Arch. Wiskd., 18, 10


\bibitem[{Meijer(1941)}]{Meijer1941}
---. 1941, Proc. Akad. Wet. Amsterdam, 44, 1062


\bibitem[{{Melnyk} {et~al.}(2006){Melnyk}, {Elyiv}, \& {Vavilova}}]{Melnyk2006}
{Melnyk}, O.~V., {Elyiv}, A.~A., \& {Vavilova}, I.~B. 2006, Kinematika i Fizika
  Nebesnykh Tel, 22, 283


\bibitem[{M{\o}ller(1989)}]{Moller1989}
M{\o}ller, J. 1989, Adv. Appl. Probab., 21, 37


\bibitem[{M{\o}ller(1994)}]{Moller1994}
---. 1994, {Lectures on Random Voronoi Tessellations.} (Lecture Notes in
  Statistics. 87) (New York: Springer-Verlag). 
  


\bibitem[{{Okabe} {et~al.}(1992){Okabe}, {Boots}, \& {Sugihara}}]{okabe}
{Okabe}, A., {Boots}, B., \& {Sugihara}, K. 1992, {Spatial tessellations.
  Concepts and Applications of Voronoi diagrams} ({Chichester, NY}:
  {Wiley})


\bibitem[{Olver {et~al.}(2010)Olver, Lozier, Boisvert, \& Clark}]{NIST2010}
Olver, F., Lozier, D., Boisvert, R., \& Clark, C.
  2010, {NIST Handbook of Mathematical Functions.} (Cambridge: {Cambridge
  University Press.})


\bibitem[{{Pan} {et~al.}(2011){Pan}, {Vogeley}, {Hoyle}, {Choi}, \&
  {Park}}]{Vogeley2011}
{Pan}, D.~C., {Vogeley}, M.~S., {Hoyle}, F., {Choi}, Y.-Y., \& {Park}, C. 2011,
  ArXiv e-prints:1103.4156


\bibitem[{{Patiri} {et~al.}(2006){Patiri}, {Betancort-Rijo}, {Prada}, {Klypin},
  \& {Gottl{\"o}ber}}]{Patiri2006}
{Patiri}, S.~G., {Betancort-Rijo}, J.~E., {Prada}, F., {Klypin}, A., \&
  {Gottl{\"o}ber}, S. 2006, \mnras, 369, 335


\bibitem[{{Pierre}(1990)}]{pierre1990}
{Pierre}, M. 1990, \aap, 229, 7


\bibitem[{{Press} {et~al.}(1992){Press}, {Teukolsky}, {Vetterling}, \&
  {Flannery}}]{press}
{Press}, W.~H., {Teukolsky}, S.~A., {Vetterling}, W.~T., \& {Flannery}, B.~P.
  1992, {Numerical Recipes in FORTRAN. The Art of Scientific Computing}
  (Cambridge: Cambridge University Press)


\bibitem[{{Schaap} \& {van de Weygaert}(2000)}]{Schaap2000}
{Schaap}, W.~E. \& {van de Weygaert}, R. 2000, \aap, 363, L29


\bibitem[{{{Sobol}, I.M.}(1967)}]{Sobol1967}
{{Sobol}, I.M.} 1967, U.S.S.R. Comput. Math. Math. Phys., 7, 86


\bibitem[{{Subba Rao} \& {Szalay}(1992)}]{Subba1992}
{Subba Rao}, M.~U. \& {Szalay}, A.~S. 1992, \apj, 391, 483


\bibitem[{{van de Weygaert }(1991)}]{Weygaert1991a}
{van de Weygaert }, R. 1991, \mnras, 249, 159


\bibitem[{{van de Weygaert}(1991)}]{Weygaert1991b}
{van de Weygaert}, R. 1991, Ph.D. thesis, University of Leiden


\bibitem[{{van de Weygaert}(1994)}]{Weygaert1994}
---. 1994, \aap, 283, 361


\bibitem[{{van de Weygaert}(2002)}]{Weygaert2002}
---. 2002, arXiv:astro-ph/0206427


\bibitem[{{van de Weygaert}(2003)}]{Weygaert2003}
---. 2003, {Statistics of Galaxy Clustering - Commentary} (Statistical
  Challenges in Astronomy), 156--186


\bibitem[{{van de Weygaert} \& {Icke}(1989)}]{Weygaert1989}
{van de Weygaert}, R. \& {Icke}, V. 1989, \aap, 213, 1


\bibitem[{{van de Weygaert} \& {Schaap}(2009)}]{Schaap2009}
{van de Weygaert}, R. \& {Schaap}, W. 2009, in {V.~J.~Martinez, E.~Saar, E.~M.~Gonzales, \& M.~J.~Pons-Borderia}, eds, Lecture Notes in Physics Vol. 665
  (Berlin: Springer-Verlag),
  pp.~291--311.


\bibitem[{{von Benda-Beckmann} \& {M{\"u}ller}(2008)}]{Benda-Beckmann2008}
{von Benda-Beckmann}, A.~M. \& {M{\"u}ller}, V. 2008, \mnras, 384, 1189


\bibitem[{{Voronoi}(1907)}]{voronoi_1907}
{Voronoi}, G. 1907, J. Reine Angew. Math., 133, 97


\bibitem[{{Voronoi}(1908)}]{voronoi}
---. 1908, J. Reine Angew. Math., 134, 198


\bibitem[{{York} {et~al.}(2000){York}, {Adelman}, {Anderson}, {Anderson},
  {Annis}, {Bahcall}, {Bakken}, \& et~al.}]{York2000}
{York}, D.~G., {Adelman}, J., {Anderson} Jr., J.~E., {Anderson}, S.~F.,
  {Annis}, J., {Bahcall}, N.~A., {Bakken}, J.~A., \& et~al. 2000, \aj, 120,
  1579


\bibitem[{{Zaninetti}(1991)}]{zaninettig}
{Zaninetti}, L. 1991, \aap, 246, 291


\bibitem[{{Zaninetti}(2006)}]{Zaninetti2006}
---. 2006, \cjaa, 6, 387


\bibitem[{{Zaninetti}(2010)}]{Zaninetti2010g}
---. 2010, Serbian Astr. Jour., 181, 19


\end{thebibliography}

\providecommand{\newblock}{}

\appendix
\section{The Meijer $G$-function  }

\label{appendixgmeijer}
In general the Meijer $G$-function
 is defined by the following Mellin--Barnes type integral
on the complex plane,
\begin{eqnarray}{cc}
G_{p,q}^{m,n}(z) &\equiv& G_{p,q}^{m,n} \left( z\left|
\begin{array}{c} (a_i)_{1}^{p} \\\\ (b_j)_{1}^{q}
\end{array}\right.\right) \equiv  G_{p,q}^{m,n} \left( z\left|
\begin{array}{cc} a_1, \dots, a_p \\\\ b_1, \dots, b_q
\end{array}\right.\right) \nonumber\\
&=& \displaystyle\frac{1}{2 \pi i} \displaystyle\int_{\mathcal L}
\frac{\displaystyle\prod_{j=1}^m \Gamma(b_j + s)
\prod_{j=1}^{n}\Gamma(1 - a_j
-s)}{\displaystyle\prod_{j=n+1}^{p}\Gamma(a_j +
s)\prod_{j=m+1}^{q} \Gamma(1 - b_j - s)} z^{-s} ds,
\label{AppG}
\end{eqnarray}
where the contour of integration $\mathcal L$ is arranged to lie
between the poles of $\Gamma(a_i+s)$ and the poles of
$\Gamma(b_j+s)$. The $G$-function is defined under the following
hypothesis.
\begin{itemize}
\item $0 \leq m \leq q, 0 \leq n \leq p$, and $p \leq q-1$; \item
$z \neq 0$; \item no pair of $b_j$, $j$ distinct, $j = 1,2,\dots,m$ differ by
an integer or zero; \item the parameters $a_i \in \mathrm{C}$
and $b_j\in \mathrm{C}$ are such that no pole of $\Gamma (b_j + s),
j = 1,2,\dots,m$ coincide with any pole of $\Gamma (a_i+s), i =
1,2,\dots,n$; \item $a_i - b_j \neq 1,2,3,\dots$ for $i =
1,2,\dots,n$ and $j = 1,2,\dots,m$; and \item if $p=q$, then the
definition makes sense only for $|z|<1$,
see \citet{Meijer1936,Meijer1941,H02007,NIST2010}.
\end{itemize}
\end{document}